\documentclass[epj]{svjour}
\usepackage{graphics}
\usepackage{color,soul}

\usepackage{hyperref}
\usepackage{cite}

\definecolor{Orange}{cmyk}{0.,0.5,0.5, 0.}

\def\order#1{\ensuremath{{\cal O}(#1)}}
\newcommand{\al}{\alpha}
\newcommand{\als}{\al_s}
\newcommand{\GaH}{\Gamma_H}
\newcommand{\MH}{M_H}
\newcommand{\mt}{m_t}
\newcommand{\mb}{m_b}
\newcommand{\lsim}
{\;\raisebox{-.3em}{$\stackrel{\displaystyle <}{\sim}$}\;}

\newcommand{\gev}{\,\, \mathrm{GeV}}
\newcommand{\mev}{\,\, \mathrm{MeV}}
\newcommand{\MZ}{M_Z}
\newcommand{\MW}{M_W}
\newcommand{\seff}[1]{\sin^2\theta_{\rm eff}^{#1}}
\newcommand{\as}{\al_s}
\newcommand{\at}{\al_t}

\begin{document}
\title{Theory requirements for SM Higgs and EW precision physics at the FCC-ee} 
\author{S.~Heinemeyer\inst{1}\inst{2}\inst{3}
\and S.~Jadach\inst{4}
\and J.~Reuter\inst{5}
}                     
%
%
\institute{IFT (UAM/CSIC), Universidad Aut\'onoma de Madrid Cantoblanco, 8048, Spain 
\and Campus of International Excellence UAM+CSIC, Cantoblanco, 28049, Madrid, Spain
\and Instituto de F\'isica de Cantabria (CSIC-UC), 39005 Santander, Spain
\and Institute of Nuclear Physics Polish Academy of Sciences, Krakow, Poland
\and DESY, Theory Group, Notkestr. 85, 22607 Hamburg, Germany
}
\date{Received: date / Revised version: date}
%
\abstract{
High precision experimental measurements of the properties of the Higgs boson at $\sim 125$~GeV as well as electroweak precision observables such as the $W$-boson mass or the effective weak leptonic mixing angle are expected at future $e^+e^-$ colliders such as the FCC-ee. This high anticipated precision has to be matched with theory predictions for the measured quantities at the  same level of accuracy. We briefly summarize the status of these predictions within the Standard Model (SM) and of the tools that are used for their determination. We outline how the theory predictions will have to be improved in order to reach the required accuracy, and also comment on
the simulation frameworks for the Higgs and EW precision program. 
\PACS{
      {PACS-key}{discribing text of that key}   \and
      {PACS-key}{discribing text of that key}
     } 
} 
\maketitle


\section{Introduction}
\label{sec:intro}


With the discovery of the Higgs boson, all possible elements of the Standard Model (SM) have
been experimentally confirmed and tested in great depth. On the other hand,
observational evidence for dark matter and the
matter-antimatter asymmetry require physics beyond the SM. One
promising way to probe such new physics is through precision measurements of the
properties of the Higgs boson. A complementary way is to measure electroweak precision (pseudo-)observables (EWPO) with highest precision. These are the
avenues pursued by several proposals for a future $e^+e^-$ collider. In
particular, the FCC-ee concept is designed to run at $\sqrt{s} = 250 \gev$ as a Higgs factory, and at $\sqrt{s} \sim \MZ, 2 \MW$ for high-precision EWPO measurements. In this way it
can improve (indirect) probes for
new physics by several orders of magnitude compared to existing bounds~\cite{Gomez-Ceballos:2013zzn,Abada:2019zxq,Abada:2019lih}.

The anticipated experimental accuracy of an observable has to be matched
with a theory prediction of at least the same level of accuracy to make maximum
use of the experimental data.
For the accurate study of the properties of the Higgs boson, precise
predictions for the various partial decay widths, the branching ratios
(BRs) and the Higgs-boson production cross sections 
along with their theoretical uncertainties are indispensable.  
For the EWPO on the one hand the SM prediction has to reach the level of the experimental accuracy. 
Similarly, the extraction of these quantities from experimental data must be equally well under control.
All theses types of uncertainties (various theoretical ones and experimental) must be taken into
account when deriving constraints on new physics from the data.

Several sources of theory uncertainties have to be distinguished. 
The {\em intrinsic} uncertainties are due to missing higher orders in
the perturbative expansion of the SM (or BSM) prediction for an observable. 
The {\em parametric} uncertainties
are due to the imperfect experimental knowledge of the SM input
parameters (as well as theory uncertainties induced in their extraction from 
data). The extraction of a quantity from a cross section or an
asymmetry requires the theory prediction of this cross section or
asymmetry to at least the same order of precision%
\footnote{Theoretical predictions for the past electron collider experiments (LEP/SLC)
were at least a factor 3 more precise than experimental data.}.
%

In this paper the current status and future implications of theory uncertainties
on (SM) Higgs-boson physics and EWPO's will be summarized. 
While we focus on the FCC-ee, they are valid for
all future high precision $e^+e^-$ colliders running above the $HZ$ threshold (such as 
ILC, CLIC, or CEPC). We will use anticipated FCC-ee precisions
to illustrate the impact of theory uncertainties. 
We build on~\cite{Freitas:2019bre}, where also a discussion of the determination of the SM input parameters can be found. In addition, we review Monte Carlo (MC) event generator frameworks which are indispensable for simulations in the perturbative and non-perturbative regime, building
upon the experience gained from the LEP era~\cite{Kobel:2000aw,Jadach:1996gu,Beenakker:1996kt} and the extensive simulations for CEPC, ILC and CLIC. Work on dedicated FCC-ee MC samples has started.


\section{Pseudo-observables versus realistic observables}
\label{sec:pseudo}

The quantities that can be directly measured in experiments are cross
sections, line shape observables, forward--backward asymmetries etc., 
called ``realistic observables'' \cite{Bardin:1999gt,ALEPH:2005ab}.
The obtained results depend on the specific set of experimental cuts,
however detector inefficiencies are removed and kinematic 
cuts are brought to simplified form using MC event generators.
In order to determine idealized quantities like 
masses, partial widths or couplings, or the effective electroweak
mixing angle, from the primarily measured
realistic observables, a ``QED unfolding'' procedure is applied. 
This procedure involves manipulations like subtracting 
photon-exchange or interference terms, subtracting box-diagram contributions, 
unfolding higher-order QED or QCD corrections, etc.
These secondary quantities are therefore called ``pseudo-observables''. 
The Higgs-boson observables and EWPOs on which we focus here are pseudo-observables.
It should be stressed that the role of EWPOs is to represent and encapsulate
experimental data in the {\em model-independent} way%
\footnote{For instance $\seff{f}$ is not a SM parameter --
it parametrizes {\em data} on the ratio of the vector to axial couplings
in $e^+e^-\to f\bar{f}$.}.
Fitting EWPOs to the SM and its extensions comes as an independent next step. 
In order to reach the required theoretical accuracy special care has to be taken in the application of these unfolding procedures -- in particular
the procedure of the extraction of EWPOs from data causes certain bias,
partial loss of information on physics in the realistic data.
The unfolding procedure for the LEP experiments 
is described in Refs.~\cite{Bardin:1999gt,ALEPH:2005ab}.
A discussion of the additional problems in this respect arising in beyond the SM (BSM) models can be found in~\cite{Heinemeyer:2004gx}. 
In Refs.~\cite{Bardin:1999gt,ALEPH:2005ab} it was proven that the bias induced
in the extraction of EWPOs (QED deconvolution) was smaller than the LEP experimental errors.
This will not be true in the FCC-ee era where experimental errors
will be smaller by up to two orders.
The very survival of the technique of the model independent EWPOs as the bridge between 
the experiment and the theory at FCC-ee will require significant improvement
of the precision of the trivial but large QED effects.
Ref.\cite{Jadach:2019bye} covers this issue in a great detail, 
listing challenges and outlining solutions.
In addition, the definition of EWPOs and the algorithm of their extraction
will have to be modified. 
This is elaborated in Sect.~3 in Ref.~\cite{Blondel:2018mad}.
Two major changes are anticipated. One is that we may be forced
to remove from EWPOs not only QED but also
first order pure electroweak corrections,
in order to achieve sharper resolution in the search of 
the BSM effects in the new EWPOs.
Another important change will be the increased role of Monte Carlo generators
in extracting EWPOs from data -- this is because semianalytic
programs like ZFITTER and TOPAZ0 of LEP era will not be able to handle
precisely enough detector geometry (cut-offs).


\section{Electroweak precision observables}
\label{sec:ewpo}

The most important electroweak precision observables (EWPOs) are related
to properties of the $Z$ and $W$ bosons. 
The $Z$-boson properties are determined from measurements of 
$e^+e^- \to f\bar{f}$ on the $Z$-pole. 
To isolate the physics of the $Z$-boson, the typical set of pseudo-observables 
is defined in terms of the de-convoluted cross-section $\sigma_f(s)$, 
where the effect of initial- and final-state photon radiation 
and from $s$-channel photon and double-boson (box) exchange  has been removed.
The customary set of pseudo-observables are total and partial $Z$-boson cross sections 
with their subsequent decay to quarks or leptons, forward-backward and left-right asymmetries,
from which the effective weak leptonic mixing angle, $\seff{f}$, is extracted, 
as well as the $W$-boson mass, $\MW$, see Ref.~\cite{Schael:2013ita}.
The expected (pure) experimental precision at FCC-ee of $\MW$ 
and $\seff{\ell}$ is summarized in Tab.~\ref{tab:ewpo-fccee}
based on Refs.~\cite{Abada:2019zxq,Abada:2019lih}.

\begin{table}[tb]
\centering
\renewcommand{\arraystretch}{1.1}
\begin{tabular}[t]{l|cclc}
\hline
Quantity & FCC-ee & \multicolumn{2}{l}{Current intrinsic unc.} & Projected intrinsic unc. \\
\hline
$\MW$ [MeV]               & 0.5  & 4    & ($\al^3, \al^2\as$)   & 1 \\
$\seff{\ell}$ [$10^{-5}$] & 0.6  & 4.5  & ($\al^3, \al^2\as)$   & 1.5 \\
\hline
\end{tabular}
\caption{Estimated precision for the direct determination of $\MW$ and $\seff{\ell}$ at
FCC-ee~\cite{Abada:2019zxq,Abada:2019lih} 
(column two), current intrinsic uncertainty (column three) and projected future uncertainty when leading 3-loop corrections become available (column 4).
}
\label{tab:ewpo-fccee}
\end{table}

The quantities listed in Tab.~\ref{tab:ewpo-fccee} can be predicted within the SM in terms of the input parameters $G_{F}$, 
$\alpha(\MZ)$, $\as(\MZ)$, $\MZ$, $\MH$ and $\mt$.  
The radiative corrections in these predictions
are currently known including complete two-loop corrections
In addition, approximate three- and
four-loop corrections of \order{\at^3}, \order{\at^2\as}, 
\order{\at\as^2} and \order{\at\as^3} are available (with $\at =
y_t^2/(4\pi)$ and $y_t$ being the top Yukawa coupling), see~\cite{Freitas:2016sty} for a review. 
The theory uncertainties from missing
higher-order corrections are given in the third column of   
Tab.~\ref{tab:ewpo-fccee}~\cite{Awramik:2003rn,Awramik:2004ge,Freitas:2014hra,Dubovyk:2018rlg}. 
Also indicated are the main sources for the respective uncertainties.
In order to match the FCC-ee precision the theory predictions will have to be improved substantially. It was estimated in~\cite{Freitas:2014owa,Freitas:2019bre,Blondel:2018mad} that the intrinsic uncertainties will be reduced by complete \order{\alpha\as^2} corrections, fermionic \order{\alpha^2\as} corrections, double-fermionic \order{\alpha^3} corrections, and leading four-loop corrections enhanced by the top Yukawa coupling, resulting in the projected intrinsic uncertainties shown in column four of Tab.~\ref{tab:ewpo-fccee}.

\smallskip
A crucial point here is that the determination of
the pseudo-observables in 
Tab.~\ref{tab:ewpo-fccee} from experimental data also requires theory input for the removal of initial-state and final-state photon radiation and $s$-channel photon exchange and box contributions. 
The theory uncertainty from missing higher QED orders is estimated to amount to a few times 0.01\% \cite{Bardin:1999gt,Jadach:2000ir,Kobel:2000aw,ALEPH:2005ab} for the $Z$-peak cross-section and total width measurements. In order not to be dominated by this uncertainty, it will  need to be reduced by about a factor 10 for the FCC-ee. This will require the
calculation of non-leading log contributions to \order{\al^3}
corrections, \order{\al^3 L^2} and \order{\al^4 L^4} contributions (with $L \equiv \log s/m_e^2$) as well as an improved treatment of fermion pair production from off-shell photons.

\smallskip
The $W$-boson mass will be determined from a threshold scan near the $W$-pair threshold, $\sqrt{s} \approx 161\gev$. It is foreseen that the experimental uncertainty at FCC-ee for this measurement is about $0.5 \mev$~\cite{Gomez-Ceballos:2013zzn,Abada:2019zxq,Abada:2019lih}. 
At the point of highest sensitivity, an uncertainty of the cross-section
measurement of $0.1\%$ translates to an uncertainty of $\sim 1.5 \mev$ on
$M_W$~\cite{Azzi:2017iih}. Therefore a theoretical prediction for the
process $e^+e^-\to 4f$ with an accuracy of $\Delta\sigma\sim 0.01\%$ is
desirable, including effects of off-shell $W$~bosons, which become
important near threshold.

The currently best calculations are based on complete one-loop results for $e^+e^- \to 4f$~\cite{Denner:2005fg} and partial higher-order effects for the total cross section from an effective field
theory framework \cite{Beneke:2007zg,Actis:2008rb}. The resulting theory
uncertainty on $\MW$ is estimated to be about 3~MeV \cite{Actis:2008rb}.
This result must be improved by complete 2-loop corrections to $e^+e^- \to W^+W^-$ and to $W \to f\bar{f}'$ (based on the effective
field theory framework). In addition, a more accurate description of
initial-state radiation will be important, which includes universal
contributions from soft and collinear photon radiation (see~\cite{Bardin:1999gt,Jadach:2000ir,Kobel:2000aw} for a review), as well as hard photon radiation. For the latter, a proper matching and
merging procedure needs to be employed to avoid double counting
\cite{Denner:2000bj,Beneke:2007zg}, see~\cite{Freitas:2019bre} for more details. There it was estimated that a theory induced systematic uncertainty of $\Delta \MW \lsim 0.60 \mev$ can be feasible.


\section{The SM Higgs boson}
\label{sec:higgs}

\subsection{Higgs-boson production}
\label{sec:prod}

The very narrow width of the Higgs boson allows for a factorization of all
cross-sections with resonant Higgs bosons into production and decay
parts to very high precision if the Higgs boson can be fully reconstructed. 
In this case, finite-width effects and off-shell
contributions are of relative size $\GaH/\MH \sim 0.00003$ and thus
not relevant. If the Higgs boson is not fully reconstructable
(e.g.\ in $H\to W^{(*)}W^{(*)} \to 2\ell2\nu$) Higgs off-shell contributions have to be
taken into account (which is straightforward at NLO).

At the FCC-ee with $\sqrt{s} = 240$~GeV (or other $e^+e^-$ machines near this
center-of-mass energy), the Higgs-boson production cross-section is strongly
dominated by $e^+e^- \to ZH$, and $e^+e^- \to \nu \bar\nu H$ contributes
less than 20\%~\cite{Gomez-Ceballos:2013zzn,Moortgat-Picka:2015yla}. 
For these two processes full one-loop corrections in the SM are
available~\cite{Belanger:2002ik,Denner:2003yg}. For the dominating $ZH$
production mode they are found at the level of $\sim 5{-}10\%$. 
Leading two-loop corrections to this cross section were evaluated at 
\order{\al\als}~\cite{Gong:2016jys,Sun:2016bel}, which were found to amount to $\sim 1-2\%$.
This number has to be compared to the anticipated experimental accuracy
of 0.4\%~\cite{Gomez-Ceballos:2013zzn,Abada:2019zxq,Abada:2019lih}. It becomes clear that with a full two-loop calculation
of $e^+e^- \to ZH$ the intrinsic uncertainty will be sufficiently
small. Calculational techniques for $2 \to 2$ processes at the two-loop
level exist, and it is reasonable to assume that, if required, this
calculation within the SM can be incorporated for the FCC-ee 
Higgs precision studies (see also~\cite{Song:2021vru} for recent progress in this direction.)

For WBF production, the calculation of the full two-loop corrections will
be significantly more difficult, since this is a $2 \to 3$ process. However, one
may assume that a partial result based on diagrams with closed light-fermion
loops and top-quark loops (in a large-$\mt$ approximation) can be achieved,
which should reduce the intrinsic theory uncertainty to below the 1\% level.
Given the fact that the WBF process is less crucial than the $HZ$ channel for
the Higgs physics program FCC-ee with $\sqrt{s} = 240$~GeV,
this will probably be adequate for most practical purposes.
It would be desirable to have a complete $2\to4$ calculation at two loops, including both processes with their interference, but very likely factorized processes with on-shell projections will be sufficient.


\subsection{Higgs-boson decay}
\label{sec:decay}

The current intrinsic and parametric uncertainties
for the various Higgs-boson decay widths are given in
Tab.~\ref{tab:Hintr-para} (see~\cite{Freitas:2019bre}). 
The status of the intrinsic uncertainties was evaluated following~\cite{Lepage:2014fla,deFlorian:2016spz,lhchxswg-br,Denner:2011mq}.

\begin{table}[tb]
\centering
\renewcommand{\arraystretch}{1.1}
\begin{tabular}[t]{l|ccc|ccc}
\hline
Partial width & intr.\ QCD & intr.\ electroweak & total & para.\ $m_q$ & para.\ $\als$ & para.\ $\MH$ \\
\hline
$H \to b \bar b$ & $\sim 0.2\%$ & ${<0.3}\%$ & ${<0.4}\%$ & ${1.4\%}$ & ${0.4\%}$ & {--} \\
$H \to c \bar c$ & $\sim 0.2\%$ & ${<0.3}\%$ & ${<0.4}\%$ & ${4.0\%}$ & ${0.4\%}$ & {--} \\
$H \to \tau^+\tau^-$ & -- & ${<0.3}\%$ & ${<0.3}\%$ & {--} & {--} & {--} \\
$H \to \mu^+\mu^-$ & -- & ${<0.3}\%$ & ${<0.3}\%$ & {--} & {--} & {--} \\
$H \to gg$ & $\sim 3\%$ & $\sim 1\%$ & $\sim 3.2\%$ & ${<0.2\%}$ & $3.7\%$ & {--} \\
$H \to \gamma\gamma$ & $< {0.1}\%$ & $< 1\%$ & ${<} 1\%$ & ${<0.2\%}$ & {--} & {--} \\
$H \to Z\gamma$ & ${\lsim 0.1}\%$ & $\sim 5\%$ & $\sim 5\%$ & {--} & {--} & ${2.1\%}$ \\
$H \to WW \to 4$f & $< 0.5\%$ & ${< 0.3\%}$ & $\sim 0.5\%$ & {--} & {--} & ${2.6\%}$ \\
$H \to ZZ \to 4$f & $< 0.5\%$ & ${< 0.3\%}$ & $\sim 0.5\%$ & {--} & {--} & ${3.0\%}$ \\
\hline
\end{tabular}
\caption{
Current intrinsinc and parametric uncertainties in the various Higgs-boson decay width
calculations, see text and Refs.~\cite{Lepage:2014fla,deFlorian:2016spz,lhchxswg-br,Denner:2011mq}. ``--'' indicates a negligible source of
uncertainty.
}
\label{tab:Hintr-para}
\end{table}

Also the parametric uncertainties can play a non-negligible role for the
evaluation of the partial widths. The most important parameters are the bottom quark mass
and the strong coupling constant. In Ref.~\cite{deFlorian:2016spz} the current
uncertainties of $\als$ and $\mb$ have been assumed to be $\delta\als = 0.0015$
and $\delta \mb = 0.03 \gev$. Additionally, $\delta m_{c} =
0.025\gev$, $\delta \mt = 0.85\gev$ and $\delta\MH = 0.24\gev$~\cite{Zyla:2020zbs} have been taken into account. The
effect on the various partial widths has been evaluated as in Ref.~\cite{Lepage:2014fla} 
and is shown in the three right columns of Tab.~\ref{tab:Hintr-para}. 

When comparing the combined intrinsic and parametric uncertainties 
with the target precision of FCC-ee~\cite{Gomez-Ceballos:2013zzn,Abada:2019zxq,Abada:2019lih}, see the last column in Tab.~\ref{tab:fcceeh}, 
it is clear that improvements are necessary.
Concerning the intrinsic theory uncertainty, the available predictions for the
$f\bar{f}$ and $\gamma\gamma$ channels are already
sufficiently precise to match the expected FCC-ee experimental uncertainty. 
With available calculational techniques, the evaluation of complete two-loop
corrections to $H\to f\bar{f}$ can be achieved. This would reduce the
uncertainty of the electroweak contributions to less than 0.1\%. Similarly, the
complete NLO corrections to $H \to Z\gamma$ can be carried out with
existing methods, resulting in an estimated precision of about 1\%.
More theoretical work is needed for $H\to WW, ZZ, gg$, which are currently
limited by QCD uncertainties. 
For $H \to gg$ the calculation of massless four-loop QCD diagrams, will be required, which may be
within reach \cite{Baikov:2005rw}, reducing the intrinsic uncertainties to the level of about 1\%.
For $H \to WW, ZZ$, the required QCD corrections are essentially
identical to those for $e^+e^- \to WW$, and 
it is straightforward to improve them to a practically negligible level.
Further significant progress would require the calculation of two-loop electroweak
corrections, which for a $1\to 4$ process is beyond reach for the forseeable
future.
Here it should be noted, however, that the $HZZ$ coupling will be mostly constrained by the
measurement of the $e^+e^- \to HZ$ production process at FCC-ee with
$\sqrt{s} = 240$~GeV, rather
than the decay $H \to ZZ^*$, see the discussion in section~\ref{sec:prod}, leading to a remaining intrinsic uncertainty of less than 0.3\%.

\begin{table}[tb]
\centering
\renewcommand{\arraystretch}{1.1}
\begin{tabular}[t]{l|c|ccc|l}
\hline
decay & projected intr.\ & para.\ $m_q$ & para.\ $\als$ & para.\ $\MH$ & FCC-ee prec.\ on ${g^2_{HXX}}$ \\
\hline
$H \to b \bar b$ & $\sim 0.2\%$ & $0.6\%$ & $<0.1\%$ & -- & {$\sim 0.8\%$} \\
$H \to c \bar c$ & $\sim 0.2\%$ & $\sim 1\%$ & $<0.1\%$ & -- & {$\sim 1.4\%$} \\
$H \to \tau^+\tau^-$ & $< 0.1\%$ & -- & -- & -- & {$\sim 1.1\%$} \\
$H \to \mu^+\mu^-$ & $< 0.1\%$  & -- & -- & -- & {$\sim 12\%$} \\
$H \to gg$ & $\sim 1\%$ &  & $0.5\%$ (0.3\%) & -- & {$\sim 1.6\%$} \\
$H \to \gamma\gamma$ & $<1\%$ & -- & -- & -- & {$\sim 3.0\%$} \\
$H \to Z\gamma$ & $\sim1\%$ & -- & -- & $\sim 0.1\%$ & \\
$H \to WW$ & $\lsim 0.3\%$ & -- & -- & $\sim 0.1\%$ & {$\sim 0.4\%$} \\
$H \to ZZ$ & $\lsim 0.3\%^\dagger$ & -- & -- & $\sim 0.1\%$ & {$\sim 0.3\%$} \\
\hline
{$\Gamma_{\rm tot}$} & {$\sim 0.3\%$} & {$\sim 0.4\%$} 
                        & {$< 0.1\%$} & {$< 0.1\%$} & {$\sim 1\%$} \\
\hline
\multicolumn{4}{l}{$^\dagger$ From $e^+e^- \to HZ$ production}
\end{tabular}
\caption{
Projected intrinsic and parametric uncertainties for the partial and
total Higgs-boson decay width predictions (see text).
The last column shows the target of FCC-ee precisions on the respective
coupling squared. 
}
\label{tab:fcceeh}
\end{table}

Also shown in Tab.~\ref{tab:fcceeh} are the projected parametric
uncertainties, assuming FCC-ee precisions.
For inputs, we use $\delta\als = 0.0002$ and $\delta \mt = 50\mev$~\footnote{This assumes a precision from an $e^+e^-$ top threshold scan. At the highest FCC-ee energy, 365~GeV, the top Yukawa coupling can also be inferred from the top threshold dependence on electroweak loop corrections.}, 
$\delta\MH \sim 10\mev$, and $\delta \mb \sim 13\mev$ and $\delta m_{\rm c} \sim 7\mev$, see~\cite{Freitas:2019bre} for details. 
%
\footnote{
Note that the numbers in Tab.~\ref{tab:fcceeh} do not take into account
correlations between the uncertainties in the Higgs production and decay
processes or between different decay processes, in particular entering
via $\Gamma_{\rm tot}$. Their impact can only be
evaluated when the full experimental correlation matrix is known, see the discussion in~\cite{Freitas:2019bre}.}
It becomes clear that the intrinsic and parametric uncertainties will be able to match the high anticipated experimental accuracy at FCC-ee.


\section{Monte-Carlo and exclusive predictions}
\label{sec:mc}

To determine systematic uncertainties, to properly extract parameters from the 
Higgs measurements and to find deviations from SM predictions, 
it is indispensable to have Monte Carlo (MC) simulations.
The challenge compared to LEP/SLC times is the strongly increased experimental precision and much higher luminosities which demands much better simulations.
The precise description of QED effects, 
both regarding the correct normalization of cross sections 
as well as exclusive multi-photon radiation will be one of the highest priorities.
In the recent years, there has been steady progress in dealing 
with the calculation of higher logarithmic and finite orders in the QED radiation, 
enhanced by the very small electron mass, see 
Refs.~\cite{Ablinger:2020qvo,Blumlein:2020jrf,Blumlein:2021jdl,Frixione:2019lga,Bertone:2019hks}. 
These very advanced calculations are only partially applicable for
the exclusive description of photon radiation which is necessary if cuts are to be applied on those photons. At the level
of LO matrix elements, there is a matching procedure between explicit matrix element photons and the electron PDFs in collinear
factorization in the context of \texttt{WHIZARD} (cf. below) which has been recently shown to work unexpectedly well~\cite{Kalinowski:2020lhp,1859727}. 

From the LEP era there are many dedicated MC programs which feature 
soft photon resummation in the exclusive form.
Most of them {\tt YFS2/KORALZ} \cite{Jadach:1999tr},
{\tt BHLUMI} \cite{Jadach:1996is}, 
{\tt BHWIDE} \cite{Jadach:1995nk}, 
{\tt YFSWW} \cite{Placzek:2003zg}
are based directly on the Yennie-Frautschi-Suura (YSF) resummation scheme~\cite{Yennie:1961ad}.
A more powerful coherent exclusive exponentiation (CEEX) 
at the amplitude level was introduced in~\cite{Jadach:2000ir} 
and is so far implemented only in \texttt{KKMC}~\cite{Jadach:1999vf} program
for two-fermion processes $e^+e^-\to f\bar{f}$, $f=\mu,\tau,\nu,q$. 
A generic semi-automatic implementation of CEEX for a wider class of processes
would be a very desirable development, and might pave the path towards a 
consistent inclusion of the non-soft higher order contributions over the complete 
photon phase space, merging together genuine electroweak corrections
with QED correction calculated to much higher orders, 
inclusion of spin correlations, initial-final state interferences and more. The ultimate goal is 
to match non-soft QED corrections up to forth order and genuine SM electroweak corrections up to  second order.

Multi-purpose event generators allow to simulate the full spectrum of SM (and BSM) processes at FCC-ee. They provide (for the moment) a lower level of 
precision for the two-fermion processes, but allow to simulate them in the same framework as four-, six- and event eight-fermion processes. Due to the much higher precision and better detectors than at SLC/LEP these processes have to be included already at the $250 \gev$ stage, and not only if triple electroweak resonances or top pairs are kinematically accessible. 
In this context a lot of experience has been gained from the large full SM event samples generated for 
ILC, CLIC and CEPC using \texttt{WHIZARD}~\cite{Moretti:2001zz,Kilian:2007gr}. 
This supports leading-logarithmic QED collinear factorization for QED photons, 
automated NLO QCD corrections for arbitrary processes, 
while NLO electroweak corrections are under way. For experimental studies, 
crucial observables are the correct numbers of neutral and charged hadrons. 
These can only be correctly simulated if the color-flow assignments are correctly made, 
which includes that the invariant masses of the shower systems from the most prominent processes, $e^+e^- \to W^+W^-,ZZ$ are preserved; in a full electroweak matrix element they are, 
however, quantum mechanically entangled.
The MC hence needs to determine the correct probability of the resonant subprocesses 
and hand them over to parton shower and hadronization~\cite{Sjostrand:2006za,Sjostrand:2014zea}. 
It is important to note that the solution of the long-standing problem of the soft photon
resummation to narrow width semi-stable charged resonances like $W^\pm$ was recently
outlined in Ref.~\cite{Jadach:2019wol}.
It is also very important to take photon-induced backgrounds from Weizs\"acker-Williams QED splittings into account, particularly $\gamma\gamma\to$~hadrons.
\texttt{WHIZARD} has a special treatment for the top threshold run~\cite{Bach:2017ggt} that is used by the FCC-ee collaboration. It also supports all different kinds of beam spectra, from simple parameterized files, over Gaussian beam spreads to a sophisticated 
beamstrahlung setup for linear and circular lepton machines~\cite{Ohl:1996fi}. 

To search for deviations in the Higgs sector from the SM, \texttt{WHIZARD} supports a large setup of BSM models like supersymmetry, extended Higgs models like 2HDM, Higgs singlet extensions, SMEFT, composite and Little Higgs models, which can be generically extended to its interface to
automated Lagrangian model tools, cf.~e.g.~\cite{Christensen:2010wz,Degrande:2011ua}


\section{Outlook and deliverables}
\label{sec:conclusion}

The most important steps to be taken in the future to match the high anticipated accuracy of FCC-ee in the context of Higgs physics and EWPO with SM theory predictions are the following.

\begin{itemize}
    \item 
    Improved unfolding techniques to go from observables to pseudo-observables.

    \item
    Calculations for the EWPO: complete \order{\alpha\as^2} corrections, fermionic \order{\alpha^2\as} corrections, double-fermionic \order{\alpha^3} corrections, and leading four-loop corrections enhanced by the top Yukawa coupling.
    
    \item
    To extract the $W$-boson mass: complete 2-loop corrections to $e^+e^- \to W^+W^-$ and to $W \to f\bar{f}'$ (based on the effective field theory framework).
    
    \item 
    Development on matching scheme between EW corrections and radiated/resummend photons.
    
    \item
    Full two-loop corrections for $e^+e^- \to ZH$ and $e^+e^- \to VVH$
    
    \item
    Full two-loop corrections for $H \to ff (+\gamma)$.
    
    \item
    (Mostly) automatized MC generation including NLO QCD and EW corrections.  
    
    \item
    Improvements by roughly one order of magnitude in the determination of the most important SM parameters ($m_t$, $m_b$, $\als$, $\Delta\al_{\rm had}$, $\MH$, \ldots)
    
    \item
    Complete simulations of relevant SM processes at different energy stages in full detector simulation
    
\end{itemize}

\bigskip
Finally, the following should be kept in mind.
The SM constitutes the model in which highest
theoretical precision for the predictions of EWPO and Higgs-boson observables can be obtained. This concerns their prediction as well as their extraction. However, as
soon as BSM physics will be discovered, an evaluation of the predictions of the 
EWPO and the Higgs-boson observables in any preferred BSM model will be necessary.
The corresponding theory uncertainties, both intrinsic and parametric,
can then be larger
(see, e.g., \cite{Freitas:2013xga,Heinemeyer:2004gx} for the Minimal Supersymmetric SM).  
A dedicated theory effort (beyond the SM) would be needed in this case.


\subsection*{Acknowledgements}

The work of S.H.\ is supported in part by the
MEINCOP Spain under contract PID2019-110058GB-C21 and in part by
the AEI through the grant IFT Centro de Excelencia Severo Ochoa SEV-2016-0597.
S.J. acknowledges funding from the European Union's Horizon 2020 research and innovation programme under under grant agreement No 951754 and support of the National Science Centre, Poland, Grant No. 2019/34/E/ST2/00457.
JRR acknowledges the support by the Deutsche Forschungsgemeinschaft (DFG, German Research Association) under Germany's Excellence Strategy-EXC 2121 "Quantum Universe"-39083330.
The authors would like to thank Tord Riemann, who passed away as this report was written, for his decade-long work on EW precision physics for $e^+e^-$ colliders.

\bibliographystyle{elsarticle-num} 

\bibliography{jst} 

\end{document}